\documentclass[aps, prb,reprint, groupedaddress, showpacs, showkeys,twocolumn]{revtex4-1}
\usepackage{graphicx}
\usepackage{amsmath, color, ulem}
\newcommand{\dpartial}[2]{\frac{\partial #1}{\partial #2}}
\newcommand{\ve}[1]{{\bf #1}}

\newcommand{\mw}[1]{\left\langle #1 \right\rangle}

\newcommand{\kb}[0]{k_{\rm B}}

\usepackage{soulutf8}%erlaubt das farbliche markieren des Textes
\usepackage{upgreek}

\begin{document}

%\preprint{}

\title{Thermally induced magnon accumulation in two-sublattice magnets}

\author{Ulrike Ritzmann}
\affiliation{Institute of Physics, Johannes Gutenberg University Mainz, D-55099 Mainz, Germany}
\author{Denise Hinzke}
\affiliation{Department of Physics, University of Konstanz, D-78457 Konstanz, Germany}
\author{Ulrich Nowak}
\affiliation{Department of Physics, University of Konstanz, D-78457 Konstanz, Germany}
%\thanks{}

\date{\today}

\begin{abstract}
We present a temperature dependent study of the thermal excitation of magnonic spin currents in two-sublattice magnetic materials. Using atomistic spin model simulations, we study the local magnetization profiles in the vicinity of a temperature step in antiferromagnets, as well as in ferrimagnets. It is shown that the strength of  the excitation of the spin currents in these systems scales with the derivative of the magnetization with respect to the temperature. 
\end{abstract}

\pacs{75.30.Ds, 75.10.Hk, 75.50Gg, 75.76.+j} 
%\keywords{}

\maketitle

\section{Introduction}
The spin Seebeck effect (SSE) \cite{Uchida_2008, Uchida2010, Uchida_2010a}, which was discovered in 2008, describes the excitation of a spin current via thermal gradients. In the presence of a temperature gradient, net magnonic spin currents propagate from the hotter towards the colder direction. These currents can be detected in a metallic contact material using spin pumping and the inverse spin Hall effect \cite{Saitoh_2006}. This new techniques opens the possibility for new design concepts of thermoelectric applications including the interplay of heat, charge and spin \cite{Bauer_2012, Boona_2014b, Uchida_2014b}.

In the last years, a variety of experimental studies have been performed to investigate the characteristics of thermally excited spin currents, as the influence of the interface \cite{Lu_2013}, the dependence on the thickness of the magnetic sample \cite{Kehlberger_2015}, and the influence of the temperature or external magnetic fields \cite{Guo_2015, Kikkawa_2013b, Kikkawa_2015, Jin_2015, Ritzmann_2015}. In recent experiments, it was shown that the SSE is significantly enhanced by using multilayer structures \cite{Ramos_2015, Lee_2015}. 

To describe these experimental findings, theoretical models have been developed \cite{Xiao_2010, Hoffman_2013, Ritzmann_2014, Kehlberger_2015, Ritzmann_2015, Etesami_2014, Rezende_2014}. These theories, based on magnonic spin currents as origin of the SSE, cover some of the observed characteristics, for example the thickness dependence of the effect. It was demonstrated experimentally that the measured spin Seebeck coefficient increases with increasing thickness for thin films and later saturates on a characteristic length scale. Due to a comparison of experiments and numerical simulation, this saturation effect can be referred to a finite magnon propagation length \cite{Kehlberger_2015}. Moreover, strong magnetic fields are used to control the SSE and suppress low frequency magnons of the excited spin currents. By performing atomistic spin model simulations, it has been shown that the contributions of the so-called subthermal magnons is much higher as expected, since the frequency dependent propagation length 
for these magnons is larger than the averaged magnon propagation length \cite{Ritzmann_2015}.  

But the mentioned descriptions base on a simplified ferromagnetic model. However, the most common material for the magnetic insulator in the SSE measurements is Yttrium iron garnet (YIG), which is a ferrimagnetic material \cite{Cherepanov_1993}. The observed complex temperature dependence of the SSE in YIG  \cite{Guo_2015}, as well as the observed magnetic field dependence indicate that the SSE depends on the frequency spectrum of the excited magnons. Therefore, a ferrimagnetic model is needed to include the complex frequency spectra of the magnons. 

First predictions for antiferromagnets and compensated ferrimagnets in a two-sublattice ferrimagnet by Ohnuma et al.~claim that even if the SSE is vanishing for antiferromagnets, the thermally excited spin current is non-vanishing for compensated ferrimagnets \cite{Ohnuma_2013}. Moreover, in recent experiments in Gadolinium iron garnet (GdIG) two sign changes of the spin Seebeck effect have been observed \cite{Gepraegs_2014}. One sign change was observed at the compensation point of the magnetization.

In this paper, we study numerically the thermal excitation of spin currents in two-sublattice materials in the vicinity of a temperature step. We show with atomistic spin model simulations that, in agreement with earlier predictions, the spin current is vanishing for degenerated antiferromagnets. In ferrimagnetic systems with compensation point of the magnetization, we investigate the temperature dependence of the excited magnon accumulation and link the results with the difference of the equilibrium magnetization of the two heat baths as driving force of the thermally excited magnon accumulation. This shows that even around the compensation point, at which the magnetization vanishes a spin current can be excited.
%the spin Seebeck effect. This dependence can be used to find new materials with high spin Seebeck coefficients in the wanted temperature range for future applications.

\section{Methods}
This work focusses on a numerical study of the thermally excited magnon accumulation in two-sublattice magnets. For that purpose, we model a simple cubic lattice with magnetic moments that can differ between the two sublattices A and B. We consider an Hamiltonian including exchange interactions with isotropic exchange coupling $J_{ij}$ between the magnetic moments $i$ and $j$ and an anisotropy with an easy-axis in $z$-direction described by the anisotropy constant $d_z$, leading to
\begin{align} 
    \label{hamiltonian}
    \mathcal{H}=-\sum_{i,j}{J_{ij}\ve{S}_i\ve{S}_j}-d_z\sum_i{\big(S_{i}^z\big)^2}\;\mbox{.}
  \end{align} 
Classical spin model simulations have been performed by solving the stochastic Landau-Lifshitz-Gilbert equation for each normalized magnetic moment $\ve{S}_i$ \cite{Nowak_2007},
  \begin{align}
    \label{LLG}
    \dpartial{\ve{S}_i}{t}=-\frac{\gamma}{\mu_{\rm s}(1+\alpha^2)} \ve{S}_i\times\left(\ve{H}_i+\alpha\left(\ve{S}_i\times\ve{H}_i\right)\right)\mbox{,}
  \end{align}  
where $\gamma$ is the gyromagnetic ratio, $\mu_{\rm s}$ the magnetic moment, $\alpha$ the phenomenological damping constant. The effective field $\ve{H}_i$ of each magnetic moment is given by  
 \begin{align}
    \label{Heff}
    \ve{H}_i=-\dpartial{\mathcal{H}}{\ve{S}_i}+\boldsymbol{\zeta}_i(t)\;\mbox{.}
  \end{align}
The temperature is included as additive white noise $\boldsymbol{\zeta}_i(t)$ to the effective field. The numerical integration of the equations of motion for each normalized magnetic moment is very time consuming, but provides a correct temperature dependence of the magnetization and the full frequency spectra of the magnons without artificial cutoffs due to discretization effects. % Quelle eventuell schonmal verwendet bei Magnetfeldpaper??!!

In a first step, an antiferromagnetic system has been studied. In this case, the interaction of only nearest neighbors is considered and the exchange constant is given by $J_{ij}=J<0$. In a second step, the model is extended to a ferrimagnetic system with two different magnetic moments of the single sublattices, including also next-nearest-neighbor interaction. In both cases, the magnetic moments are initialized parallel or antiparallel to the easy-axis in $z$-direction. In the considered system, a time independent temperature profile is applied, describing the lattice temperature of the systems. This profile includes a temperature step in the center along the longest axis of the system from the temperature $T_1$ to $T_2$. 

Due to this temperature step, the magnetic system is in a non-equilibrium state. The system relaxes and reaches a quasistationary state, in which the local magnetization and the excited spin current are on average time-independent. Since in the hotter region more magnons exist than in the colder one, a net magnon current from the hotter towards the colder region of the system is created and causes deviations of the local magnetization from the equilibrium value, as studied earlier for a ferromagnetic system \cite{Ritzmann_2014}. Possible reflections at the cold end of the system, that could appear due to finite size effects are suppressed by using absorbing boundary conditions implemented by an exponential increase of the damping constant close to the end of the system. In the following the resulting magnetization profiles in the different two-sublattice magnets are discussed.

\section{Magnon accumulation in antiferromagnets}
As a first step, the thermally driven non-equilibrium states in antiferromagnets have been investigated. In the considered antiferromagnet, the magnetization of the system is vanishing, since the magnetizations of the two sublattices compensate each other at every temperature. Therefore, a thermally excited magnon current would carry no spin and no magnon accumulation would be observed. 

In oder to visualize the excited spin currents, the local magnetization in the quasi-stationary states are calculated as the average of the normalized magnetic moment $\ve{S}_i(t)$ over a time interval and over the $x$-$y$-plane for each sublattice. Similar to the ferromagnetic case studied earlier \cite{Ritzmann_2014}, the local magnetizations of the single sublattices deviate from their equilibrium values $m_0^{\mathrm{A,B}}$ close to the temperature step. These deviations are defined as magnon accumulation, given by
    \begin{align}
    \label{eq1}
     \Delta m^{A,B}(z)=m^{A,B}(z)-m_0^{A,B}(z)\textrm{.}
    \end{align}
\begin{figure}[t!]
 \includegraphics[width=0.9\columnwidth]{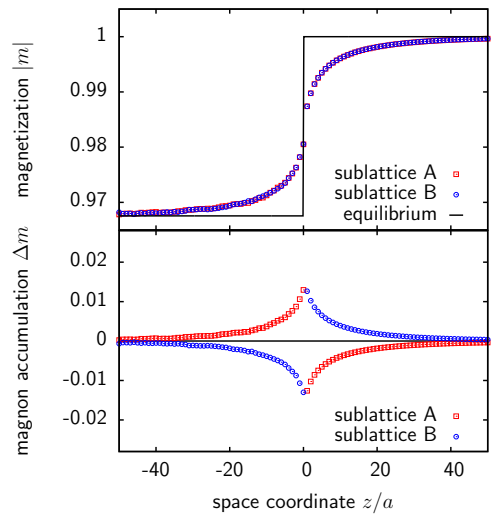}
 \caption{Absolute value of the normalized magnetization $m$ and the magnon accumulation $\Delta m$ versus the space coordinate $z$ of the single sublattices of an antiferromagnet around a temperature step exemplary for a damping constant of $\alpha=0.01$, and an anisotropy constant of $d_z=0.01J$.}
 \label{fig1}
\end{figure}
The results are shown exemplary in figure \ref{fig1} for a system with $8\times8\times512$ magnetic moments with an anisotropy constant $d_z=0.01|J|$ and a temperature step from $k_{\mathrm B}T_1=0.1|J|$ to $k_{\mathrm{B}}T_2=0$. The used damping constant is $\alpha=0.01$. The absolute value of the local magnetization $m^{A,B}(z)$ of the single sublattices is increased in the hotter region (decreased in the colder one) due to a lack (surplus) of magnons caused by the net magnonic current. This is illustrated in the upper part of figure \ref{fig1}.

In the figure below, the corresponding magnon accumulation for the sublattices, as defined in equation (\ref{eq1}), is shown. The accumulation appears only close to the temperature step and vanishes with increasing distance from the temperature step. The magnon accumulation of the single sublattices have an opposite sign and additionally, a sign change appears at the temperature step. Nevertheless, the total magnon accumulation in the antiferromagnet is totally compensated, as proposed by Ohnuma et al. \cite{Ohnuma_2013}.

\begin{figure}[t!]
 \includegraphics[width=0.99\columnwidth]{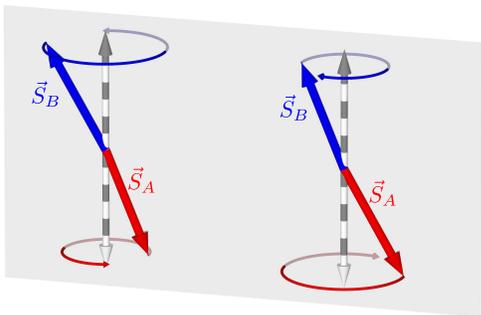}
 \caption{Illustration of the two magnon modes in an antiferromagnet with degenerated dispersion relation. The magnetic moments of the single sublattices can either precess left or righthanded leading to magnon modes transporting opposite angular momentum.}
 \label{fig1b}
\end{figure}
In a two-sublattice magnet with a cubic lattice, two magnon branches exist. Both branches can have different frequencies. For the two different modes, the precession of the single magnetic moments in the system are in opposite direction. Due to reversed amplitude ratios of the sublattices, the two modes carry angular momentum with different sign. In the absence of a magnetic field, the two magnon modes in the considered antiferromagnet are degenerated but carry opposite angular momentum of the same absolute value. The two modes are illustrated in figure \ref{fig1b}. The single magnetic moments can precess in both directions around their effective field. Due to the effective torques acting on the magnetic moments, the amplitude of one sublattice is larger than that of the other leading to a transport of angular momentum due to the magnon modes. Magnons from the two branches carry angular momentum with opposite sign. Due to their degeneracy, the magnon modes are excited thermally with the same probability. Hence, the magnon current excited in a temperature gradient do on average not transport angular momentum and the resulting total magnon accumulation vanishes. 

Nevertheless, the thermally excited magnon current, which is visible in the single sublattices, transfers heat from the hotter towards the colder region, dependent on the damping in the system. These excited magnon currents can be used for heat transfer in the system or can be used to drive a domain wall \cite{Kim_2014, Tveten_2014}. Recently is has been shown, that the thermally driven domain wall motion in antiferromagnets can be much faster than in antiferromagnetic systems \cite{Severin_2016}.

\section{Magnon accumulation in ferrimagnets}
In a next step, the thermal excitation of spin currents in a two-sublattice ferrimagnet with magnetic compensation point is investigated. Predictions by Ohnuma et al.~show a non-vanishing spin current at the compensation point \cite{Ohnuma_2013}, whereas in recent measurements by Gepr\"ags et al.~a sign change of the spin Seebeck coefficient in Gadolinium Iron Garnet was observed \cite{Gepraegs_2014}. By studying the frequency spectra in thermal equilibrium at different temperatures, the authors explain this sign change by a compensation of the involved magnon modes.

In this paper, we consider a minimal model for a ferrimagnet including magnetic compensation. In this system, the second sublattice B has a larger magnetic moment than the sublattice A, but demagnetizes at lower temperatures due to the additional next-nearest neighbor interaction. We have chosen a lower intra-sublattice exchange interaction for the second sublattice, $J_B<J_A$. In particular, we have chosen a ratio of the magnetic moments of the sublattices of $\mu_{\mathrm B}=1.5\mu_{\mathrm A}$, and exchange constants of $J_B=0.2J_A$ and $J=-0.1J_A$, where the dominant exchange interaction term is given by the interaction of the magnetic moments of sublattice $A$ \cite{Schlickeiser_2012}. 

\begin{figure}[t!]
 \includegraphics[width=0.9\columnwidth]{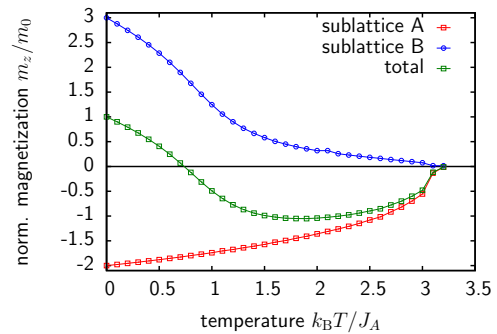}
 \caption{Normalized magnetization component $m_z/m_0$ of the single sublattices A and B, and of the total system versus temperature $T$ in thermal equilibrium.}
 \label{fig2}
\end{figure}

The magnetization $\ve m$ is given as the averaged magnetic moment over the whole system. It can be written as $\ve{m}=\mu_A\mw{\ve{S}_A}+\mu_B\mw{\ve{S}_B}$. At a temperature of $\kb T=0$, the magnetization is given by $|m_0|=|\mu_A-\mu_B|=0.5\mu_A$. The temperature dependence of the normalized equilibrium magnetization for this system is illustrated in figure \ref{fig2}. In this figure, only the $z$-component of the averaged magnetization of the whole system is shown, aligned with the easy axis of the model. The other two components vanish on average. Using the $z$-component of the magnetization instead of its absolute value includes additionally the information about the direction of the magnetization. This plays an important role for the determination of the magnon accumulation around the compensation point, where 
the direction of the magnetization 
changes. 

As shown in figure \ref{fig2}, the magnetization at low temperatures is aligned in the direction of the magnetization of sublattice $B$, since its magnetic moment is larger. But due to the lower exchange interaction of sublattice B, it demagnetizes at lower temperatures and the magnetization is compensated at a temperature  of $k_{\mathrm B}T_{\mathrm{comp}}=0.7J_A$ changing its direction. Finally, the magnetization vanishes at a critical temperature of $k_{\mathrm B}T_{\mathrm C}=3.2J_A$.

In this system, we have investigated the thermally driven spin currents in the vicinity of a temperature step dependent on the temperature level of the system. To resolve the magnon accumulation at high temperatures, we have used a system consisting of $512\times32\times32$ magnetic moments. This larger system leads to a huge computational effort of the performed simulations, but it enables us to average out thermal fluctuations and to detect local magnetization profiles with a high accuracy even at high temperatures. Moreover, we have calculated the average over four simulations to resolve even small magnon accumulations. This allows us to identify a sign change of the thermally created magnon accumulation at high temperatures. 

\begin{figure}[t!]
 \includegraphics[width=0.9\columnwidth]{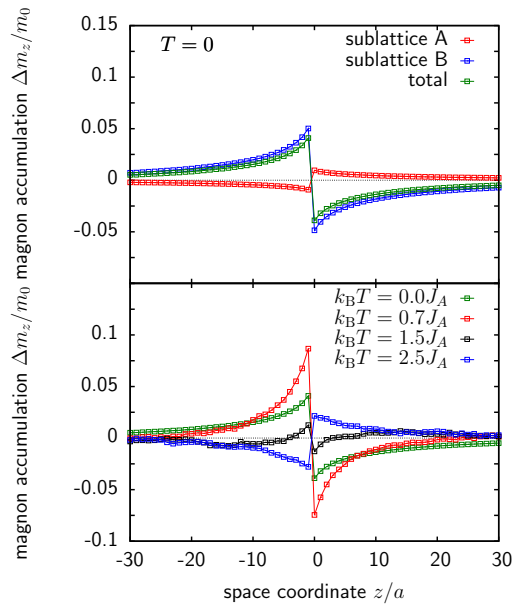}
 \caption{Magnon accumulation due to a temperature step for different temperature levels. Top: Magnon accumulation $\Delta m_z/m_0$ versus the space coordinate $z$ of the single sublattices and the total magnon accumulation due to a temperature step. Bottom: Total magnon accumulation $\Delta m_z/m_0$ versus the space coordinate $z$ for different temperatures $T$.}
 \label{fig3}
\end{figure}

We have studied, similar to the antiferromagnetic case, a system with a constant temperature step at the center of the $z$-direction with a temperature difference of $k_{\mathrm B}\Delta T=0.1 J_A$ and a damping constant of $\alpha=0.01$. Additionally, we have calculated the magnon accumulation due to this temperature step for different temperature levels. The resulting magnon accumulation in the quasistationary state of the system is shown exemplary for $k_{\mathrm B}T_2=0$ in the upper part of figure \ref{fig3}. As before, a net magnon current appearing due to the temperature gradient creates deviations from the equilibrium value. In the hotter region the number of magnons is reduced in comparison to the equilibrium situation and, therefore, the absolute value of the magnetization of the single sublattices is increased. In the cold region the situation is reversed and the additional magnons cause a reduction of the absolute value of the sublattice magnetizations. 

This leads to the shown magnon accumulation of the single sublattices. In contrast to antiferromagnets, the total magnon accumulation, which is given as the sum over the contributions of the two sublattices, is no longer compensated. At low temperatures, it is dominated by the change of the magnetization of sublattice B.
By varying the used temperature level of the temperature profile, the magnon accumulation shows a strong temperature dependence. This is illustrated for different temperatures in the lower part of figure \ref{fig3}. In the low temperature regime, the magnon accumulation increases at first and shows a maximum close at the compensation point. In agreement with the predictions by Ohnuma et al. \cite{Ohnuma_2013}, the magnon accumulation does not change its sign at the compensation point,  $k_{\mathrm B}T_{\mathrm{comp}}\approx0.7J_A$. For even higher temperatures, the magnon accumulation decreases and shows a sign change around a temperature of $k_{\mathrm B}T=1.8J_A$. The accumulation vanishes at the critical temperature.

Also in the two-sublattice ferrimagnet that is used, two magnon branches exist. In contrast to the antiferromagnet, the branches are not degenerated and they are thermally excited with different probabilities. The modes affect the magnetization of both sublattices, but dependent on the mode, the amplitude of each mode in one sublattice is larger than in the other defining the polarization of the mode. The lower frequency mode carries angular momentum proportional to $-m_z\ve{e}_z$. The higher mode carries the opposite angular momentum. In this case, the amplitude of the second sublattice is more affected. 

The thermally excited spin current is given by the sum of the magnons from both branches. The strength of the magnon current of each branch scales in first order with the difference of the magnon densities in the two heat baths. At low temperature, the resulting magnon accumulation is dominated by the contribution of the low frequency branch. An increase of the magnon accumulation should therefore be due to an increase of the difference of the magnon densities of the low frequency branch. At higher temperatures, the higher branch becomes more important and the transported spin current is reduced since more and more magnons with opposite spin polarization are contributing to the spin current. At a characteristic temperature the contributions from the two branches cancel each other and the magnon accumulation vanishes. Above this temperature, the higher branch is dominating and the spin current has a opposite sign.

This described temperature dependence of the magnon accumulation can be linked on a macroscopic scale with the change of the magnetization between the two heat baths. If the magnetization varies a lot with temperature, the magnon density of one or both magnon branches varies a lot. Therefore, one can assume that the magnon accumulation scales with the magnetization slope in thermal equilibrium. 

\begin{figure}[t!]
 \includegraphics[width=0.9\columnwidth]{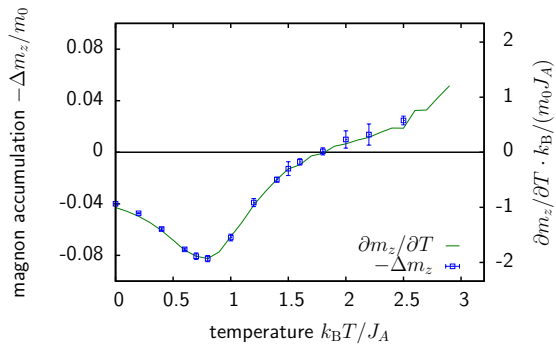}
 \caption{Strength of the magnon accumulation $\Delta m_z(T)$ at the temperature step as a functon of the temperature in comparison with the derivative of the magnetization with respect to the temperature $\partial m/\partial T$ as a function of the temperature $T$.}
 \label{slope}
\end{figure}

In order to derive the gradient of the magnetization with respect to the temperature, $\partial m_z/\partial T$, numerically for the studied system, the difference quotient of the magnetization due to a temperature difference $\Delta T=0.1J_A$ is calculated up to the second order. To compare this slope with the magnon accumulation, the strength of the magnon accumulation directly at the temperature step is considered, which is defined by
\begin{align}
     \Delta m_z(T)=\frac{1}{2}\big(\Delta m_z(-a,T+\Delta T)-\Delta m_z(0,T)\big)
\end{align}
The calculated slope of the magnetization as well as the magnon accumulation are shown in figure \ref{slope}. The temperature dependence of both quantities show a very good agreement over the whole temperature range. The magnon accumulation at the temperature step scales linear with the gradient of the magnetization with respect to the temperature with a scaling factor of $\Delta m(0)\approx -23.4 \;J_A/k_{\mathrm B} \cdot \partial M/\partial T$. This scaling factor depends on system parameters as the damping constant $\alpha$, since for lower damping more magnons propagate around the interface. Hence the strength of the magnon accumulation at the temperature step is increased for lower damping.

This scaling explains the non-vanishing magnon accumulation around the compensation point. Since the first sublattice demagnetizes at low temperatures, the magnetization varies a lot with temperature. This is still true at the compensation point and, therefore, the magnon accumulation is around its maximum value. At higher temperatures, sublattice B is mainly demagnetized and the magnetization changes less with temperature and the strength of the magnon accumulation decreases. Around a temperature of $k_{\mathrm{B}}T=1.8J_A$, the magnetization does only weakly vary with temperature and the gradient as well as the magnon accumulation vanishes. At even higher temperatures, the magnetization increases with temperature, leading to a gradient and a magnon accumulation with opposite sign. 

\begin{figure}[t!]
 \includegraphics[width=0.9\columnwidth]{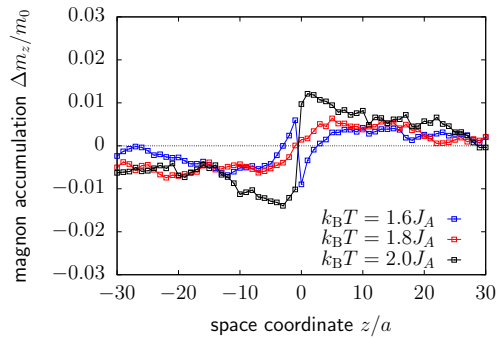}
 \caption{Magnon accumulation due to a temperature step as a function of the position $z$ for different temperature levels close to the compensation point of magnon accumulation.}
 \label{sign}
\end{figure}

Furthermore, we observe a sign change of the magnon accumulation at a temperature of $k_{\mathrm B}T=1.5J_A$ with increasing distance to the temperature step, as shown in figure \ref{fig3}. %This behavior indicates a strong dependence of the magnon accumulation on the propagation length of the single magnon modes.} 
%Dependent on the distance from the source of the spin current, the sign of the spin current can change. These effect could be measured for example in the so-called local spin Seebeck geometry, that was studied recently for YIG \cite{Cornelissen_2015, Goennenwein_2015, Cornelissen_2016}.} 
In figure \ref{sign}, the magnon accumulation for temperature levels around $k_{\mathrm{B}}T=1.8J_A$ are shown. According to the results shown in figure \ref{slope}, the magnon accumulation directly at the interface changes its sign around this temperature. Nevertheless, the sign of the magnon accumulation far away from the temperature step is the same below and above the sign change of the magnetization slope. 
This can be explained by the different length scales for magnon propagation of the two different modes. If the involved magnons from the higher frequency branch have a higher propagation length than the magnons from the lower one, the relative contribution of the higher branch increases with distance to the temperature step and can even dominate. Hence, the sign of the magnon accumulation can change. The observations of this effect shows that the magnon modes have different length scales. This could be measured, for example, in the so-called non-local spin Seebeck geometry that was studied recently for YIG \cite{Cornelissen_2015, Goennenwein_2015, Cornelissen_2016}.

\section{Summary}
In this paper, we present a study of the excitation of thermally induced spin currents in two-sublattice materials by investigating the magnon accumulation that is induced due to a temperature step. We show in agreement with earlier predictions by Ohnuma et al.\cite{Ohnuma_2013} that in antiferromagnets the magnon accumulation and the corresponding spin current vanish. Nevertheless, in a compensated two-sublattice ferrimagnet the thermally induced magnon accumulation does not vanish around the compensation point.

The observed magnon accumulation scales with the change of the magnetization with respect to a temperature change. By studying the temperature dependence of the magnon accumulation, we show a correlation between the slope of the magnetization and the strength of the magnon accumulation. Therefore, the maximum of the magnon accumulation appears at the inflection point of the magnetization curve, whereas the signal vanishes if the magnetization does not change with temperature.

This macroscopic link demonstrates that thermally excited spin currents are caused by a gradient in the magnon densities. In agreement with results in ferromagnets, also in ferrimagnets the gradient of the magnon densities cause a magnonic spin current. But in two-sublattice materials, both magnons branches have to be included, leading to a complex temperature dependence of the driving force of thermally induced spin currents. Using these results, one can identify new materials with high spin Seebeck coefficients in the preferred temperature range, by simply studying the temperature dependence of the magnetization.

\begin{acknowledgments}
 The authors would like to thank the Deutsche Forschungsgemeinschaft (DFG) for financial support via SPP 1538 ``Spin Caloric Transport'' and the SFB 767 ``Controlled Nanosystem: Interaction and Interfacing to the Macroscale''.
\end{acknowledgments}

\bibliography{./Quellen}

\end{document}